\documentclass[twocolumn,prl,showpacs,floatfix]{revtex4}

\usepackage{graphicx}
\usepackage{bm}
\usepackage{color}
\usepackage{amsmath}
\usepackage{amsfonts}
\usepackage{amssymb}
\usepackage{epstopdf}
\usepackage{ulem}

\begin{document}

\preprint{APS/123-QED}

\title{Localized magnetic excitations in the fully frustrated dimerized magnet Ba$_2$CoSi$_2$O$_6$Cl$_2$}

\author{Nobuyuki Kurita$^1$, Daisuke Yamamoto$^2$, Takuya Kanesaka$^2$, Nobuo Furukawa$^2$, Seiko Ohira-Kawamura$^3$, Kenji Nakajima$^3$, and Hidekazu Tanaka$^1$}
\affiliation{
$^1$Department of Physics, Tokyo Institute of Technology, Oh-okayama, Meguro-ku, Tokyo 152-8551, Japan\\
$^2$Department of Physics and Mathematics, Aoyama-Gakuin University, Sagamihara, Kanagawa 252-5258, Japan\\
$^3$Materials and Life Science Division, J-PARC Center, Tokai, Ibaraki 319-1195, Japan
}
\date{\today}

\begin{abstract}
Magnetic excitations of the effective spin $S$\,=\,1/2 dimerized magnet Ba$_2$CoSi$_2$O$_6$Cl$_2$ have been probed directly via inelastic neutron scattering experiments at temperatures down to 4\,K.
We observed five types of excitation at 4.8, 5.8, 6.6, 11.4, and 14.0\,meV, which are all dispersionless within the resolution limits.
The scattering intensities of the three low-lying excitations were found to exhibit different $\bm{Q}$-dependences.
Detailed analysis has demonstrated that Ba$_2$CoSi$_2$O$_6$Cl$_2$ is a two-dimensional spin dimer system described only by a single dimer site,
where the triplet excitations are localized owing to the almost perfect frustration of the interdimer exchange interactions and 
the undimerized spins, even in small concentration, make an essential contribution to the excitation spectrum.

\end{abstract}



\maketitle


A coupled spin dimer system, in which antiferromagnetic (AF) dimers interact with one another through interdimer exchange interactions, provides an opportunity for correlating condensed matter physics with particle physics. 
One intriguing feature is the crystallization of magnetic quasiparticles, magnons (or triplons), like a Wigner crystal, which could be a key to understanding exotic quantum phases 
such as supersolids\,~\cite{Chen_PRB2010,Albuquerque_PRB2011,Murakami_PRB2013,Yamamoto_PRB2013} and flat-band solid states\,~\cite{Richter_PRB2018}.
The crystallized phase of magnons is expected to emerge under a magnetic field
 when the frustration of the interdimer exchange interactions is
so strong that magnons become localized\,~\cite{Kageyama_PRL1999,Miyahara_PRL1999,Tanaka_JPSJ2014}.
This quantum phenomenon can be characterized by a stepwise magnetization process and dispersionless magnetic excitations.
Until recently, experimental studies have been, to our best knowledge, 
limited to the Shastry--Sutherland compound SrCu$_2$(BO$_3$)$_2$\,\cite{Kageyama_PRL1999,Miyahara_PRL1999,Onizuka_JPSJ2000,Momoi_PRB2000,Kodama_Science2002,Miyahara_JPCM2003,Takigawa_PRL2013,Matsuda_PRL2013}. In this compound, magnons are localized owing to the orthogonal configuration of dimers.
Fractional magnetization plateaus observed in SrCu$_2$(BO$_3$)$_2$ imply the successive crystallization of magnons.

\begin{figure}
\begin{center}
\includegraphics[width=0.9\linewidth]{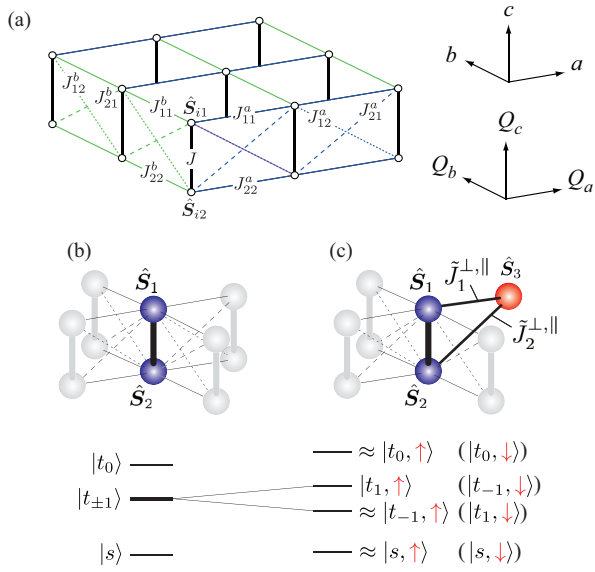}
\end{center}
\caption{(Color online) 
Schematic views of (a) 2D exchange network considered for Ba$_2$CoSi$_2$O$_6$Cl$_2$
and emergent (b) two-body and (c) three-body problems.
Real (upper) and reciprocal (lower) space coordinations are shown on the right of (a).
The lower diagrams show the energy levels with the corresponding eigenstates,
where the energy levels of the three-body problem are all doubly degenerate 
with the spin-inversion counterparts of each eigenstate shown in parentheses.
} \label{fig1}
\end{figure}

The ground state of a spin dimer system is typically a spin singlet with an excitation gap $\Delta$ to the lowest excited triplet state.
When a magnetic field exceeding the critical field $H_{\rm c}$(=$\Delta/g\mu_{\rm B}$) is applied, 
magnons are created on the dimer lattice\,~\cite{Rice_Science2002}.
Magnons can hop to neighboring dimer sites and interact with each other via transverse and longitudinal components of interdimer exchange interactions, respectively.
For the simplified two-dimensional (2D) case, as illustrated in Fig.~\ref{fig1}(a), the hopping and repulsive terms are proportional to $(J_{11}$\,+\,$J_{22})$\,$-$\,$(J_{12}$\,$+$\,$J_{21})$ and  $J_{11}$\,+\,$J_{22}$\,$+$\,$J_{12}$\,$+$\,$J_{21}$, respectively.
Given that the frustration of interdimer exchange interactions is perfect, 
namely $J_{11}$\,+\,$J_{22}$\,=\,$J_{12}$\,$+$\,$J_{21}$,
the hopping of magnons becomes completely suppressed and magnons
form a periodic array consisting of half-filled magnons 
owing to the competition between the repulsive interactions of magnons and the Zeeman energy.
When the hopping term is dominant, by contrast, the spin dimer system undergoes an XY-type AF ordering upon applying a magnetic field of above $H_{\rm c}$.
It is known that a magnetic-field-induced quantum phase transition can be described by the Bose--Einstein condensation (BEC) of magnons\,\cite{Affleck_PRB1991,Giamarchi_PRB1999,Nikuni_PRL2000}.
Magnon BEC has been experimentally verified by comparative measurements 
using many spin dimer compounds such as TlCuCl$_3$ and BaCuSi$_2$O$_6$\,~\cite{Ruegg_Nature2003,Yamada_PRB2008,Jaime_PRL2004,Zapf_PRL2006,Review}.

Recently, the magnetic insulators Ba$_2$CoSi$_2$O$_6$Cl$_2$~\cite{Tanaka_JPSJ2014} and Ba$_2$CuSi$_2$O$_6$Cl$_2$~\cite{Okada_PRB16} were reported to be a new series of 2D spin dimer systems with the exchange network shown in Fig.~\ref{fig1}(a).
High field magnetization measurements of  Ba$_2$CoSi$_2$O$_6$Cl$_2$ up to 70\,T
revealed the complete magnetization process with a magnetization plateau 
at half of saturation magnetization $M_{\rm s}$\,\cite{Tanaka_JPSJ2014}.
While the edges of the magnetization plateau reported for SrCu$_2$(BO$_3$)$_2$ are rather smeared,
the (1/2)$M_{\rm s}$ plateau observed in Ba$_2$CoSi$_2$O$_6$Cl$_2$ is sharply stepwise. 
This could suggest that interdimer exchange interactions are almost perfectly frustrated.
On the other hand, a spin dimer system can give rise to a similar (1/2)$M_{\rm s}$ plateau provided that there exist two kinds of isolated dimer with equal populations.
For a definitive conclusion, it is important to elucidate the magnetic excitations of this compound.
Ba$_2$CoSi$_2$O$_6$Cl$_2$ crystallizes in a monoclinic structure with the space group $P2_1/c$.
The lattice parameters are $a$\,=\,7.1382\,\AA, $b$\,=\,7.1217\,\AA, $c$\,=\,18.6752\,\AA, and $\beta$\,=\,91.417$^{\circ}$.
Owing to the strong spin orbit coupling and pyramid-like crystal field, the effective spin of magnetic Co$^{2+}$ ions
can be described by an $S$\,=\,1/2 strongly XY-like XXZ model at temperatures much lower than the spin-orbit coupling constant of $|\lambda|/k_B$\,$\sim$\,250\,K.

Single crystals of Ba$_2$CoSi$_2$O$_6$Cl$_2$ were grown by a flux technique.
The detailed procedure of the crystal growth is described in the Supplemental Material (SM) of Ref.\,\onlinecite{Tanaka_JPSJ2014}.
Magnetic excitations were investigated by inelastic neutron scattering experiments using the cold-neutron disk chopper spectrometer AMATERAS 
installed in the Materials and Life Science Experimental Facility (MLF) 
at J-PARC, Japan\,\cite{Nakajima_JPSJSB11}. 
The measurements with two sets of incident neutron
energies $E_{\rm i}\,{=}$\,(2.6, 5.9, 10.5, 23.6)\,meV and (2.9, 4.7, 7.7, 15.2)\,meV were performed at several temperatures of 4\,K to 240\,K.
Approximately 60 plate-like single crystals with a total mass of $\sim$\,1\,g were glued
on an aluminum plate, where the $a$ axis (or $b$ axis) for each crystal was aligned parallel to the horizontal direction.
Note that the single crystals used in this study were twinned, where the $a$ and $b$ axes were interchanged.
The fluoropolymer (CYTOP$\textregistered$\,\cite{CYTOP})  employed as the glue had a negligible contribution to the background.
The wave vector $k_i$ of an incident neutron was set parallel to the $c^*$ axis.
All the data were analyzed using the software suite Utsusemi\,\cite{Inamura_JPSJ13}. 

\begin{figure*}[t]
\begin{center}
\includegraphics[width=0.95\linewidth]{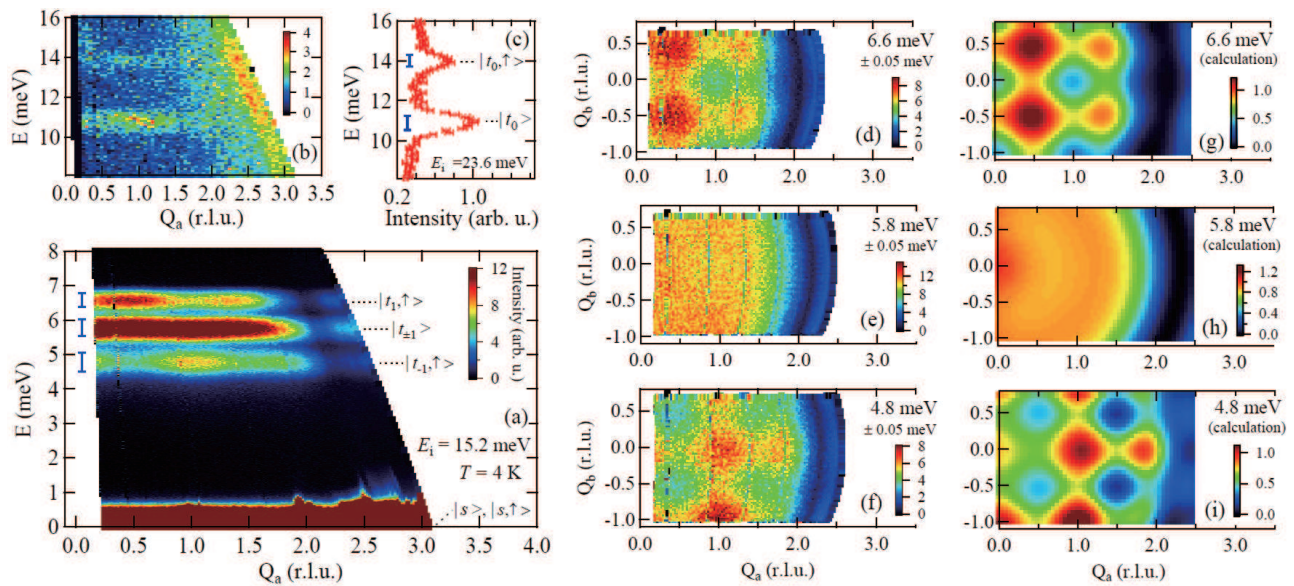}
\end{center}
\caption{(Color online)  Energy-momentum maps of the scattering intensity along ${\bm Q}\,{=}\,(Q_a, 0, 0)$ at the base temperature of 4\,K measured with an incident neutron energy of (a) $E_{\rm i}\,{=}$\,15.2\,meV and (b) 23.6\,meV, which were integrated over $Q_b$ and $Q_c$. (c) Energy vs scattering intensity measured with $E_{\rm i}\,{=}\,$23.6\,meV, where the scattering intensity was integrated over the complete $Q_a, Q_b,$ and $Q_c$ range. (d)$-$(f) Constant-energy slices of the scattering intensity, where the scattering intensity was integrated over $Q_c$, and (g)$-$(i) the calculated ones (arb. units).  In (a) and (b), vertical bars denote the energy resolution. 
}
\label{fig2}
\end{figure*}

\begin{figure}
\begin{center}
\includegraphics[width=0.95\linewidth]{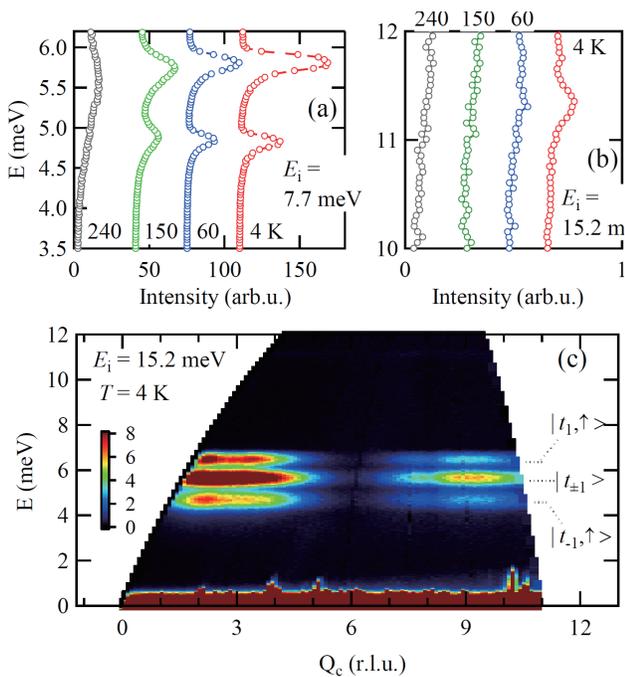}
\end{center}
\caption{(Color online) 
Temperature evolution of intensity peaks measured with $E_{\rm i}$\,=\,(a) 7.7\,meV and (b) 15.2\,meV. The scattering intensities were integrated over whole $Q_{\rm a}$, $Q_{\rm b}$, and $Q_{\rm c}$ range.
(c) Energy-momentum map of the scattering intensity along  $Q_{\rm c}$ measured with $E_{\rm i}$\,=\,15.2\,meV. The scattering intensities were integrated over $Q_{\rm a}$ and $Q_{\rm b}$.
}
\label{fig3}
\end{figure}


Figures~\ref{fig2} (a) and (b) show the energy-momentum map of the scattering intensity along ${\bm Q}\,{=}\,(Q_a, 0, 0)$ at the base temperature of 4\,K measured with an incident neutron energy of $E_{\rm i}\,{=}\,15.2$ and 23.6\,meV, respectively, where the scattering intensities were integrated over $Q_b$ and $Q_c$. 
One can confirm three strong dispersionless excitations at 4.8, 5.8, and 6.6~meV and two weak dispersionless excitations at higher energies of 11.4 and 14.0~meV.
Figure~\ref{fig2}(c) shows the scattering intensity as a function of energy measured at 4\,K with $E_{\rm i}\,{=}$\,23.6\,meV, where scattering intensities were integrated over the complete $Q_a, Q_b,$ and $Q_c$ range. 
In these figures, vertical bars denote the energy resolution.
Since the width of the excitations at 4.8, 5.8, 6.6, 11.4, and 14.0\,meV are resolution-limited, 
all the excitation peaks are single peaks and not the superposition of two or more excitation peaks.
Figures\,\ref{fig3} (a) and (b) show temperature evolutions of intensity peaks measured with $E_{\rm i}\,{=}$\,7.7 and 15.2\,meV, respectively.
With increasing temperature above 150 K, the excitation spectrum shown in Fig.\ \ref{fig2}(a) is considerably smeared and the intensity decreases while an excitation peak at 11.4\,meV is no longer detectable at $T\ge$\,60\,K. From these results, the origin was verified to be magnetic.
Additional information on the energy-momentum maps of scattering intensities for different $E_{\rm i}$ and temperatures is presented in Supplemental Material (SM)~\cite{SM} (see Figs.~SM1 and~SM2).

As shown in Fig.~\ref{fig3}(c), the scattering intensity oscillates along $Q_c$. This oscillation is related to the spin separation ${\bm R}$ in the dimer. The dynamical structure factor $S({\bm Q}, {\omega})$ is proportional to $\sin^2\{({\bm Q}{\cdot}{\bm R})/2\}$. Because in Ba$_2$CoSi$_2$O$_6$Cl$_2$, the spin separation ${\bm R}$ is approximately parallel to the $c$ direction, the oscillation of intensity occurs along $Q_c$ but not along $Q_a$ and $Q_b$. The wave vector $Q_{\rm c}^{\rm max}$ that gives the local maxima of the intensity which is proportional to $S({\bm Q}, {\omega})f^2({\bm Q})$, where $f({\bm Q})$ is the magnetic form factor of Co$^{2+}$, is calculated to be $Q_c^{\rm max}\,{=}\,3.0, 9.0\,\cdots$ (see Fig.\,SM3 in Ref.~\onlinecite{SM}). These values are consistent with the experimental results shown in Fig.~\ref{fig3}(c). Note that the decrease in scattering intensities for $1.8\,{<}\,Q_a\,{<}\,2.5$ in Fig.~\ref{fig2}(a) is caused by  the neutron absorption mainly owing to the plate-shaped samples and the sample holder of the aluminum plate, and thus is extrinsic.

The noteworthy feature of the three low-energy excitations at 4--7\,meV is that the scattering intensities exhibit different $\bm{Q}$-dependences.
This is more evident in the constant-energy slices of the scattering intensity shown in Figs.\,~\ref{fig2}(d)$-$(f), where the scattering intensity is integrated over $Q_c$, considering good two-dimensionality as evidenced by the observed dispersionless excitations along $Q_c$ [Fig.~\ref{fig3}(c)]. 
The intensity of the middle excitation with the highest intensity at 5.8\,meV is nearly independent of ($Q_a$, $Q_b$).
On the other hand, the intensities of the upper-side (6.6\,meV) and lower-side (4.8\,meV) excitations exhibit local maxima 
when both $Q_a$ and $Q_b$ are integers and half-integers, respectively.
For the excitations at 11.4 and 14.0\,meV, the low intensities make it difficult to discern their $\bm{Q}$-dependence (see Fig.\,SM4 in Ref.~\onlinecite{SM}) 
The energies of the single singlet-triplet excitation to the $|t_{\pm1}\rangle$ and $|t_0\rangle$ states are given by $E_1\,{=}\,(J^{\perp}+J^{\parallel})/2$ and $E_2\,{=}\,J^{\perp}$, respectively. 
Here, $J^{\perp}$ and $J^{\parallel}$ are transverse and longitudinal components of intradimer exchange interactions, respectively.
In Ba$_2$CoSi$_2$O$_6$Cl$_2$, the energy level of $|t_0\rangle$ is higher than that of $|t_{\pm1}\rangle$ owing to the strong XY anisotropy \cite{Tanaka_JPSJ2014}. 
It is considered that when an excited triplet is localized, the single single singlet-triplet excitation is dispersionless and its intensity is independent of $Q_a$ and $Q_b$.
Thus, the middle excitation peak at 5.8\,meV can be assigned to the single single singlet-triplet excitation to the $|t_{\pm1}\rangle$ state. Assuming that the excitation at 11.4\,meV corresponds to the single single singlet-triplet excitation to the $|t_0\rangle$ state, we obtain $J^{\perp}\,{=}\,11.4$~meV and $J^{\parallel}\,{=}\,0.16$~meV\,\cite{Anisotropy}. 
This means that the intradimer exchange interaction closely approximates the XY model.

Because there is no other singlet-triplet excitation to the $|t_{\pm1}\rangle$ states, we can deduce that all the dimers are magnetically equivalent. Thus, the sharply stepwise magnetization process with a 1/2-magnetization plateau can only be described in terms of the crystallization of localized magnons owing to the strong frustration of interdimer exchange interactions~\cite{Tanaka_JPSJ2014}.
From these observations, we can safely conclude that the interdimer interactions in Ba$_2$CoSi$_2$O$_6$Cl$_2$ almost satisfy the perfect frustration condition $J_{11}^{{\perp},{\parallel}}\,{+}\,J_{22}^{{\perp},{\parallel}}\,{=}\,J_{12}^{{\perp},{\parallel}}\,{+}\,J_{21}^{{\perp},{\parallel}}$. Note that the sharp side peaks observed at 4.8 and 6.6\,meV does not indicate the presence of the multiple dimer sites with different magnitudes of the intradimer exchange interaction $J^{{\perp},{\parallel}}$, because their intensities exhibit different periodicities that are commensurate with $a^*$ and $b^*$. 
As shown below, these side peaks rather support the perfect frustration scenario with a single dimer site.

Here we discuss the origin of the anomalous side peaks observed at 4.8 and 6.6\,meV. 
Given that the Ba$_2$CoSi$_2$O$_6$Cl$_2$ crystals employed in this study are perfect crystals, the side-peak structure will be absent from the excitation spectrum. It is natural to assume that these side peaks are produced by interdimer interactions, because the energy difference between these side peaks and the middle peak is on the order of the interdimer interactions and the intensities of these side peaks are commensurate with $a^*$ and $b^*$.

One plausible scenario is that the observed side peaks are caused by the three-body problem among dimer spins and a neighboring undimerized single spin produced by a vacancy of Co$^{2+}$, as illustrated in Fig.\,~\ref{fig1}(c).
The model Hamiltonian of the three-body problem is written as 
\begin{eqnarray}
{\cal{H}}_{\rm 3b}{\hspace{-1.5mm}}&=&{\hspace{-1.5mm}}J^\perp({S}_1^x{S}_2^x+{S}_1^y{S}_2^y)+J^\parallel{S}_1^z{S}_2^z\nonumber\\
{\hspace{-1.5mm}}&+&{\hspace{-1.5mm}}\tilde{J}^{\perp}_1({S}_1^x{S}_3^x+{S}_1^y{S}_3^y)+\tilde{J}^{\parallel}_1 {S}_1^z{S}_3^z\nonumber\\
{\hspace{-1.5mm}}&+&{\hspace{-1.5mm}}\tilde{J}^{\perp}_2({S}_2^x{S}_3^x+{S}_2^y{S}_3^y)+\tilde{J}^{\parallel}_2 {S}_2^z{S}_3^z,
\label{model3body}
\end{eqnarray}
where $\tilde{J}^{\perp,\parallel}_1$ ($\tilde{J}^{\perp,\parallel}_2$) is the exchange interaction between $\hat{\bm{S}}_{1}$ ($\hat{\bm{S}}_{2}$) and the neighboring undimerized spin $\hat{\bm{S}}_3$. $\tilde{J}^{\perp,\parallel}_1$ and $\tilde{J}^{\perp,\parallel}_2$ are in general different from the coupling constatans of the interdimer interactions in the host system and not identical to each other, $\tilde{J}^{\perp,\parallel}_1\neq \tilde{J}^{\perp,\parallel}_2$. The eigenvalues and eigenstates of the Hamiltonian (\ref{model3body}) can be easily obtained from analytical diagonalization. We set the coupling constants to $(\tilde{J}_1^{\perp}, \tilde{J}_1^{\parallel}, \tilde{J}_2^{\perp}, \tilde{J}_2^{\parallel})\,{=}\,(7.93,0.66,3.08,0.25)$ meV so that the excitation energies 4.8, 6.6, and 14.0\,meV are reproduced. Since the number of parameters is larger than the number of conditions, we used here the naive assumption that the exchange anisotropy in $\tilde{J}^{\perp,\parallel}_1$ is identical to that in $\tilde{J}^{\perp,\parallel}_2$ in order to fix the remaining degree of freedom. The discussion below is qualitatively independent of this simplification.

The dominant components of each eigenstate of the three-body Hamiltonian (\ref{model3body}) are shown in Fig. 1(c). If the undimerized spin $\hat{\bm{S}}_3$ is in the spin-up (spin-down) state, the energy of the excitation from $s$ to $t_{1}$ ($s$ to $t_{-1}$) in the neighboring dimer is higher due to the antiferromagnetic interactions while the $s$-to-$t_{-1}$ ($s$-to-$t_{1}$) excitation energy is lower, compared to the decoupled-dimer excitation energy of the host system with perfect frustration. This explains the reason why the side peaks with periodic intensity oscillation (at 4.8 and 6.6\,meV) are located above and below the middle peak (at 5.8\,meV).

To confirm the above hypothesis based on the existence of undimerized spins and to explain the periodic intensity oscillations in the excitations at 4.8 and 6.6\,meV, we calculate the differential cross section of the inelastic neutron scattering for the decoupled-dimer excitation and the excitations mediated by undimerized spins. For a system with discrete energy levels, the partial differential cross section for the transition from the ground state $|\psi_{\rm g}\rangle$ to the $n$-th excited state $|\psi_n\rangle$ is given by
\begin{eqnarray}
\frac{d\sigma^{({\rm g}\rightarrow{\rm n})}}{d\Omega}\propto\!\!\!\sum_{\alpha,\beta=x,y,z}\left(\delta_{\alpha\beta}-\frac{Q_\alpha Q_\beta}{|\bm{Q}|^2}\right)f^2(\bm{Q})S_{{\alpha}{\beta}}^{({\rm g}\rightarrow n)}(\bm{Q})
\end{eqnarray}
with the exclusive structure factor tensor
\begin{eqnarray}
S_{{\alpha}{\beta}}^{({\rm g}\rightarrow n)}(\bm{Q})&=&\sum_{ij}e^{i \bm{Q}\cdot(\bm{r}_i-\bm{r}_j)}\langle \psi_{\rm g}|\hat{S}_i^{\alpha}|\psi_n\rangle\langle \psi_n|\hat{S}_j^{\beta}|\psi_{\rm g}\rangle.\nonumber \label{Saa}
\end{eqnarray}
The excitations in the three-body problem of a dimer and its neighboring undimerized spin are characterized by the diagonal components
\begin{eqnarray}
S_{{\alpha}{\alpha}}^{({\rm g}\rightarrow n)}(\bm{Q})&=& A_n^\alpha+B_n^\alpha \cos \bm{Q}\cdot \bm{r}_{12} +C_n^\alpha \cos \bm{Q}\cdot \bm{r}_{31} \nonumber\\
&&+D_n^\alpha \cos \bm{Q}\cdot \bm{r}_{32}. \label{ABCD}
\end{eqnarray}
The relative coordinates among the three spins are given, e.g., by $\hat{\bm{r}}_{12}=(0,0,d)$, $\hat{\bm{r}}_{31}=(a,0,0)$, and $\hat{\bm{r}}_{32}=(a,0,d)$ for the case illustrated in Fig.~\ref{fig1}(c). Here, $d=|{\bm R}|=3.098\,\AA$ is the spin separation length in the dimer. The values of $A_n^\alpha$, $B_n^\alpha$, $C_n^\alpha$, and $D_n^\alpha$, which can be easily calculated by diagonalizing Eq.~(\ref{model3body}), are presented in SM~\cite{SM}. The contributions from the off-diagonal components are cancelled out by each other. Here, the slight difference between the direction of ${\bm R}$ and the crystal $c$ direction is not taken into account for simplicity. 

For the comparison with the experimental data, we consider eight different cases of the relative position between the dimer and the undimerized spin, and take the average of the contributions from the eight cases. We also perform a similar (and simpler) calculation on an isolated dimer of two spins for the single singlet-triplet excitations in the host system.
As shown in Figs.~\ref{fig2}(d)$-$(i), the calculated results give excellent agreement with the observed scattering intensities.
The intensity oscillation of the excitations at 4.8 and 6.6\,meV in the $Q_a$-$Q_b$ plane stems from the terms with $C_n^\alpha$ and $D_n^\alpha$ in Eq.~(\ref{ABCD}). We also find that a crucial factor to produce the $\bm Q$-dependent oscillation is weak mixing of singlet and triplet components in the ground and excited states owing to the entanglement with the undimerized spin state since $C_n^\alpha=D_n^\alpha=0$ if it does not occur (see SM~\cite{SM} for more details). Note that although the other set of the spin-exchange parameters obtained by interchanging $\tilde{J}^{\perp,\parallel}_{1}$ and $\tilde{J}^{\perp,\parallel}_{2}$, {i.\,e.}, $(\tilde{J}_1^{\perp}, \tilde{J}_1^{\parallel}, \tilde{J}_2^{\perp}, \tilde{J}_2^{\parallel})\,{=}\,(3.08,0.25, 7.93,0.66)$ meV gives the same excitation energies, only the case of $\tilde{J}^{\perp,\parallel}_{1}>\tilde{J}^{\perp,\parallel}_{2}$ can reproduce the correct $\bm Q$ dependence of the observed scattering intensities.

The exicitations at 4.8 and 6.6\,meV have a relatively large scattering intensity even though it is expected that there are only small amounts of undimerized spins in the crystal. The ratio among the integrated scattering intensities for the excitations at 4.8, 5.8, and 6.6\,meV is roughly estimated to be $1:1.6:1$. This can be explained from the fact that one undimerized spin affects the local excitations of its four neighboring dimers. We conclude that even a reasonably small concentration of undimerized spins ($x\approx 6$ percent from $(\frac{100-x}{2}-4x)/4x=0.8$) can provide the large intensity at 4.8 and 6.6\,meV. Note that, in the above estimation, we took into consideration the double degeneracy of the middle excitation band.

Another mechanism that might give rise to satellite peaks is the formation of a dynamic boundary state with the help of the interdimer interactions.
For instance, a single single singlet-triplet transition to the $|t_0\rangle$ state occurs first and, before the triplet excitation is relaxed to the $s\rangle$ state, the neighboring dimer is excited to the $|t_{+1}\rangle$ states to form a bound state.
Thus far, however, we could not construct specific models to reproduce the observed excitation spectra\,\cite{another}.

To conclude, we have probed the magnetic excitations of Ba$_2$CoSi$_2$O$_6$Cl$_2$ directly via inelastic neutron scattering measurements.
The five observed types of magnetic excitation are dispersionless within the resolution limits,  
and hence triplet excitations are verified to be localized.
Unexpectedly, three low-energy excitations at 4--7\,meV exhibit characteristic $\bm Q$-dependences of scattering intensities.
It was found that the excitation spectra can reasonably be explained by considering two mechanisms independently:  a ``perfect frustration'' scenario for interdimer interactions in the host system, and emergent three-body quantum states owing to undimerized spins induced by vacancies in the crystals.
This work shows that the highly frustrated quantum magnets provide the various playgrounds of interacting quantum particles, and shows a typical case in which small amount of vacancy has a large effect on the excitation spectra, although the vacancy effect is usually hidden by the spectra of the host system.
To obtain further experimental findings to increase understanding of this system, it is important to elucidate the field evolution of each excitation by, for example, in-field neutron scattering and electron spin resonance experiments.

\section*{Acknowledgments}
We thank T. J. Sato and T. Kikuchi for their technical support in the inelastic neutron scattering experiments. 
The neutron scattering experiment at J-PARC was performed under the J-PARC user program (Proposal No. 2015A0161). This work was supported by Grants-in-Aid for Scientific Research (A) (Nos.~26247058 and 17H01142), (B) (No.~17H02926), and (C) (Nos.~16K05414 and 18K03525) from Japan Society for the Promotion of Science.  


\end{document}